\documentclass{elsart}

\usepackage{latexsym}
\usepackage{epsfig}
\usepackage{amssymb}
\usepackage{graphicx}
\usepackage{dcolumn}

\newcommand{\beq}{\begin{eqnarray}}
\newcommand{\eeq}{\end{eqnarray}}
\newcommand{\be}{\begin{eqnarray*}}
\newcommand{\ee}{\end{eqnarray*}}

\newcommand{\Pom}{{\hspace{ -0.1em}I\hspace{-0.25em}P}}
\newcommand{\vphot}{{\gamma^*}}

\newcommand{\MQQ}{M_{\scriptscriptstyle{Q\bar{Q}}}}
\newcommand{\QQ}{\scriptscriptstyle{Q \bar{Q}}}

\def\lsim{\raise0.3ex\hbox{$<$\kern-0.75em\raise-1.1ex\hbox{$\sim$}}}
\def\gsim{\raise0.3ex\hbox{$>$\kern-0.75em\raise-1.1ex\hbox{$\sim$}}}

\begin{document}
\begin{frontmatter}
  \title{Coherent J/$\psi$ production \\ - a novel feature at LHC?}

  \author[oslo]{I.C. Arsene}
  \author[oslo,msu]{L. Bravina}
  \author[itep]{A.B. Kaidalov}
  \author[oslo]{K. Tywoniuk}
  \author[oslo,msu]{E. Zabrodin} 
  
  \address[oslo]{Department of Physics, University of Oslo, N-0316
    Oslo, Norway}
  \address[msu]{Skobeltzyn Institute for Nuclear Physics, Moscow State
    University, RU-119899 Moscow, Russia}
  \address[itep]{Institute of Theoretical and Experimental Physics,
    RU-117259 Moscow, Russia}
  
  \begin{abstract}
    Energy dependence of heavy quarkonia production in hadron-nucleus
    collisions is studied in the framework of the Glauber-Gribov theory.
    We emphasize a change in the space-time picture of heavy-quark state
    production on nuclei with energy. Longitudinally ordered scattering of
    a heavy-quark system takes place at low energies, while with
    increasing energy it transforms to a coherent scattering of projectile
    partons on the nuclear target. The characteristic energy scale for
    this transition depends on masses and rapidities of produced
    particles. For $J/\psi$, produced in the central
    rapidity region, the transition happens at RHIC energies. The
    parameter-free calculation of $J/\psi$ in {\it dAu} collisions is in good
    agreement with recent RHIC data. We use
    distributions of gluons in nuclei to predict suppression of heavy
    quarkonia at LHC.
  \end{abstract}
  
  \begin{keyword}
    $J/\psi$ absorption in nuclear matter \sep nuclear
    effects in hadron-nucleus collisions
    
    \PACS 13.85.-t \sep 24.85.+p \sep 25.75.-q \sep 25.75.Dw
  \end{keyword}
\end{frontmatter}

\section{Introduction}
The heavy-ion programme at RHIC (BNL) and LHC (CERN) aims at discovering
features of a possible new state of deconfined QCD matter anticipated
to form in nucleus-nucleus collisions. An important signal of the
 quark-gluon plasma (QGP) formation would be a suppression
of charmonium yield \cite{Matsui86} produced in these collisions. A
proper baseline for the discovery of this effect is charmonium production
in hadron-nucleus collisions, where the QGP is absent and only cold nuclear
matter effects are present \cite{Kharzeev97,Vogt05}.

Nuclear effects in hadron-nucleus collisions are usually discussed in
terms of the power-law parameterization
\beq
\label{eq:param}
\frac{\mbox{d}\sigma^a_{hA}}{\mbox{d}^3p} \;=\; \frac{\mbox{d}
  \sigma^a_{hN}}{\mbox{d}^3 p} \, A^{\alpha (x_F)} \;,
\eeq
where $\sigma^a_{hA}$ ($\sigma^a_{hN}$) is the inclusive cross
section of particle $a$ off a nucleus
  (nucleon). The function $\alpha(x_F)$
characterizes nuclear effects at different longitudinal momentum
fractions of the produced particle, $x_F$. For $J/\psi$ production,
measurements show a decrease of $\alpha$
from 0.95 at $x_F \approx 0$ to values $\sim 0.75$ at $x_F
\simeq 0.8$ \cite{Ald91,Lei00,Sha02} thus indicating an increase of
absorption as $x_F$ increases. Also, data
over a large range of energies show an approximate
scaling of $\alpha$ with $x_F$ \cite{Ald91,Lei00,Sha02} rather than a
scaling with $x_2$ (fraction of the total momentum carried by a parton
from a nucleus) expected from QCD factorization. 

Recently, the
substantial decrease of nuclear absorption in
$J/\psi$ production in deuteron-gold ({\it dAu}) collisions at RHIC
energy, $\sqrt{s} = $ 200 GeV, where $\sigma_{abs} \sim 1-2$ mb
\cite{PHENIX06}, compared to $\sigma_{abs} \sim 4$ mb \cite{NA5003}
measured in proton-lead ({\it pPb}) collisions at SPS
energy, $\sqrt{s} = $ 17.3 GeV has attracted a lot of attention. This
corresponds to a value of $\alpha$ consistent with 1 at $x_F = 0$.
It was widely believed that absorptive effects would increase or, at
least, remain constant with rising collision energy
\cite{Braun98,Kopeliovich01,CYR04}. In the model of \cite{CYR04}
(Sec.~4.1), e.g., the postulated growth of $\sigma_{abs}$ with energy is
motivated by a growth of charmonium-nucleon center-of-mass energy,
reflecting the rapid growth of partons carrying small momentum
fractions in the nucleon.

An equally important implication resides
in the fact that the RHIC data does not scale in $x_F$: whereas the
lower-energy data points display a flat behavior at small $x_F$,
the new points delineate a steep tilt. This novel feature seems also
to be hard to reproduce in models describing the energy dependence of
$\alpha$ \cite{Vogt00,Salgado01}.

Such a behavior of $\alpha(x_F)$ allows for a natural explanation in the
Glauber-Gribov theory of multi-particle production on nuclei
\cite{Gribov69}. At very high energies
Abramovsky-Gribov-Kancheli (AGK) cutting rules \cite{Abramovsky73}
lead to a cancellation
of the Glauber-type diagrams in the central rapidity region, i.e. for
$x_F \approx 0$, and only
so-called ``enhanced'' diagrams \cite{Kancheli70,Mueller70},
corresponding to multi-Pomeron interactions, contribute
to a difference of $\alpha$
from unity. For light quarks, this coherent
hadroproduction sets in at a typical
energy scale $E_0 \sim m_N \mu R_A$, where $R_A$ is the radius of
the nucleus, $m_N$ is the mass of a nucleon and $\mu$ is a typical
hadronic scale of the order of $\sim1$ GeV. For heavy quark states, the mass
$\MQQ$ of the heavy system introduces a new scale
\beq
\label{eq:scale}
s_M \;=\; \frac{\MQQ^2}{x_+} \, \frac{R_A m_N}{\sqrt{3}} \;,
\eeq
where $x_+ = \frac{1}{2}( \sqrt{x_F^2 + 4\MQQ^2/s} + x_F)$ is the
longitudinal momentum fraction of the heavy system.

The AGK cutting rules are violated at larger values of $x_F$ and at
low energies; the first
effect is interpreted as conservation of energy-momentum \cite{Cap76}.
The latter 
effect is in turn related to a change of the space-time picture of the
interaction \cite{Bor91}. At energies below $s_M$ longitudinally
ordered
rescatterings of the heavy system take place. In this situation we can
unambiguously define the production point of the heavy system and, in
turn, the distance it has to travel through the surrounding nuclear
matter. This leads naturally to the notion of an absorptive cross
section. At $s > s_M$ the heavy state in the projectile, which
also includes light degrees of freedom, scatters coherently off the
nucleons of a nucleus, and the conventional treatment of nuclear
absorption is not adequate. 

In the central rapidity region, the values of $s_M$
for $J/\psi$ are within the RHIC
energy range. Accordingly, the effects of shadowing of
nuclear partons become important and can be calculated using the
Glauber-Gribov theory of nuclear structure functions in the region of
$x_2 < (m_N R_A)^{-1}$.

\section{Model description and comparison to data}
\begin{figure}[t!]
  \begin{centering}
  \begin{minipage}[t]{0.5\linewidth}
    \centering
    \epsfig{figure=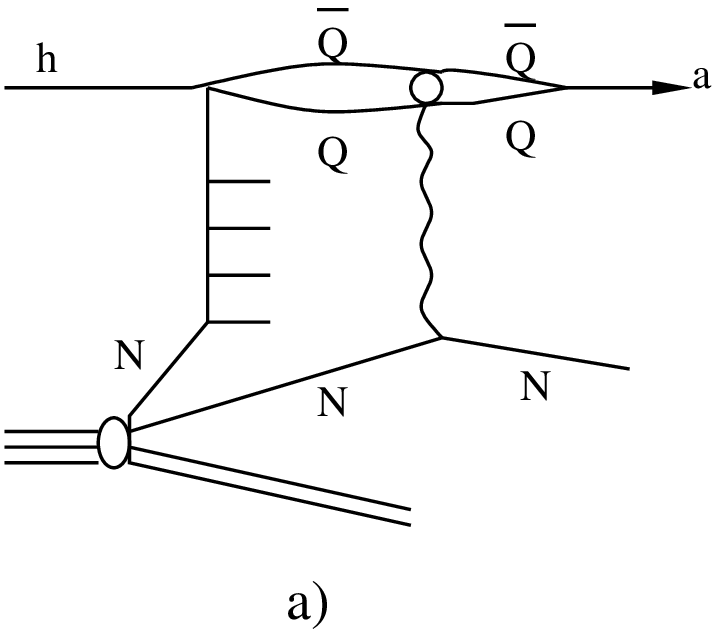,width=4cm}
  \end{minipage}%
  \begin{minipage}[t]{0.5\linewidth}
    \centering
    \epsfig{figure=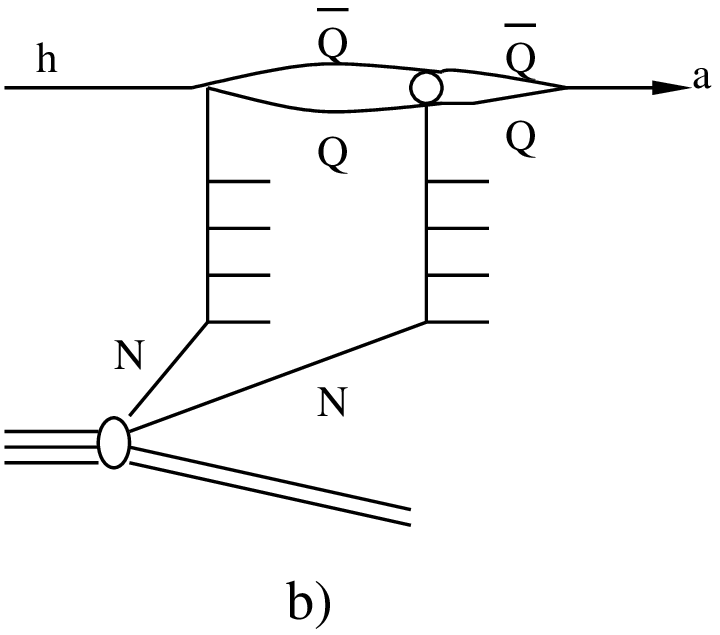,width=4cm}
  \end{minipage}\\
  \end{centering}
  \caption{\label{fig:DiagRescat} (a) Elastic and (b) inelastic
    scattering of a  $Q\bar{Q}$ system. The former does not have to be
    the leading one.}
\end{figure}

Consider first production of heavy onia in the ``low'' energy regime,
$s < s_M$. It was shown in \cite{Bor91} that the
contribution of all diagrams with intermediate heavy-quark state to
the total cross section is canceled in this energy region. However,
the different s-channel discontinuities (cuttings) of these diagrams
are different from zero. Consider, as an example, the cuttings shown
in Fig.~\ref{fig:DiagRescat}. Their contributions to the total cross section
are equal in magnitude and have opposite signs. Their contributions to
the inclusive cross
section would also cancel each other if the $Q\bar{Q}$ final states in
Fig.~\ref{fig:DiagRescat}(a) and Fig.~\ref{fig:DiagRescat}(b) would have
the same distribution in $x_F$. However, the elastic rescattering in
Fig.~\ref{fig:DiagRescat}(a) does not change the momentum, while in
the case of inelastic interaction of the $Q\bar{Q}$ system,
Fig.~\ref{fig:DiagRescat}(b), it can loose some of its momentum and
(or) transform into another state which is weakly coupled to the
observed particle $a$ (absorption). We parameterize the probability to
produce these states in a single rescattering by a parameter $(1-\epsilon)$.

Thus at energies $s < s_M$ the absorption is determined by
\beq
\label{eq:Abs}
f_{hA}^a \left( x_+ \right) = f_{hN}^a \int \mbox{d}^2b \, \frac{1 -
  e^{- \xi(x_+) \, \sigma_{\QQ} T_A (b)} }{\xi (x_+) \, \sigma_{\QQ} T_A(b)}.
\eeq
The function $f_{hN}^a (x_+) = \sigma_{hN}^a F_1^a (x_+)$, where
$F_1^a$ denotes the unmodified distribution of produced particles, and
$\xi (x_+) \;=\; (1-\epsilon) + \epsilon x^\gamma_+$ 
determines the $x_+$ dependence of the strength of the shadowing.
Finally, $T_A (b) = \int_{-\infty}^\infty \mbox{d}z \, \rho_A (b,z)$ is the
nuclear thickness function normalized to A. With $\xi (x_F =
0) = (1-\epsilon)$ one recovers the well-known Glauber formula
\cite{Ger92,Cap88} with $\sigma_{abs} = \left( 1 - \epsilon 
\right) \sigma_{\QQ}$. As shown in Ref.~\cite{Bor91},
Eq.~(\ref{eq:Abs}), with the inclusion of shadowing effects
(see below), gives a good
description of experimental data on charmonium production in pA
collisions at $E_{LAB} \, \lsim$ 800 GeV/c with $\sigma_{\QQ} = 20$ mb and
$\epsilon = 0.75$. This corresponds to an absorption cross section of
$\sigma_{abs} = 5$ mb. Note, that $\sigma_{\QQ}$ is rather large,
indicating that the $c\bar{c}$ pair is produced in the color octet
state rather than in the colorless state. It can
also be viewed as a $D\bar{D}$ ($D^*\bar{D}^*)$ system.

Equation (\ref{eq:Abs}) is not applicable at
asymptotic energies as the assumption of longitudinal
ordering is only valid at $s < s_M$. For
energies higher than $s_M$ the
expression will change due to the correct treatment of coherence
effects according to \cite{Braun98}
\beq
\label{eq:transition}
\frac{1 \,-\, e^{-\xi (x_+)\,\sigma_{\QQ} T_A(b)}}{\xi
  (x_+) \, \sigma_{\QQ}} \;\rightarrow \; T_A (b) \, e^{-
  \tilde{\sigma}_{\QQ}(x_+) T_A(b) /2} ,
\eeq
which is similar to the energy-momentum conservation effect
for light quarks \cite{Bor91}. In the model proposed in \cite{Braun98},
$\tilde{\sigma}_{\QQ}$ is equal to the total cross section of
the $Q\bar{Q} - N$ process, and is not proportional to $x_+^\gamma$.

We would like to point out that this leads to an unnatural behavior
at high energies due to the smallness of the Pomeron vertex and to an
effective double-counting of nuclear effects, and
propose an alternative procedure.
If one considers
non-enhanced Glauber-type diagrams, then the effective cross section
varies as $\tilde{\sigma}_{\QQ} \sim x_+^\gamma$, thus satisfying the
AGK cancellation. The suppression is concentrated at much 
higher $x_F$ for $Q\bar{Q}$ production than for the light hadrons because
of the large mass of the $Q\bar{Q}$ system.
It was shown in Ref.~\cite{Bor91} that at $x_F \sim 1$ the second
rescatterings in the low and high-energy limits should coincide. This
means that $\tilde{\sigma}_{\QQ} \approx \epsilon x_+^\gamma
\sigma_{\QQ}$. Experiment on $J/\psi$ production in {\it dAu}
collisions at RHIC \cite{PHENIX06} was performed in the central
rapidity region, where $x_+$ varies from 0.025 to 0.05 and
  $\tilde{\sigma}_{\QQ}$ is, therefore,
very small. This suggests an analysis of $J/\psi$ suppression in {\it
  dAu} collisions at RHIC energies taking into account enhanced
diagrams only.
A similar approach, albeit with a simpler
parameterization of nuclear shadowing, has been considered in
\cite{Cap06}.
\begin{figure}[t!]
  \centering
  \epsfig{figure=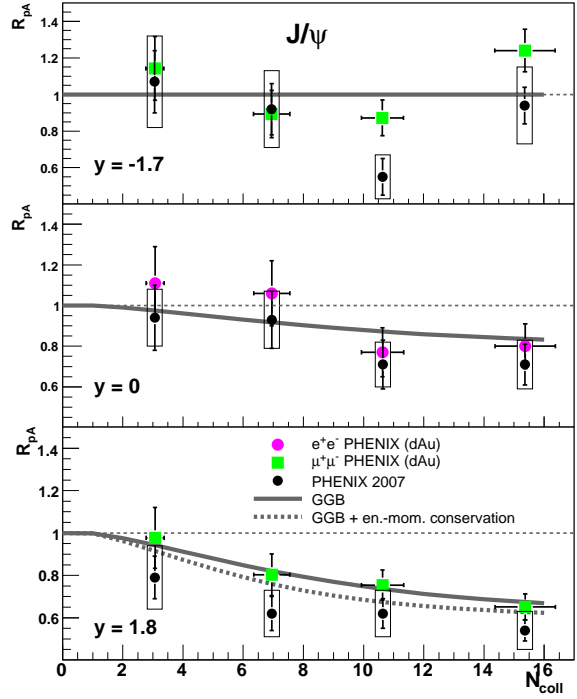,width=8.5cm}
  \caption{\label{fig:Jpsisupp}
    $J/\psi$ suppression as a function of centrality at
    different rapidities at RHIC. Data are taken from
    \cite{PHENIX06,PHENIX07}.}
\end{figure}

We have studied gluon shadowing in
Ref.~\cite{Tywoniuk05,Tywoniuk06,Ars07}, where a model for $\vphot$A
collisions was
considered within the Glauber-Gribov theory \cite{Gribov69}
including enhanced diagrams or, in other words, interactions among Pomerons.
Summing up an arbitrary number of Pomeron tree diagrams as in the
generalized Schwimmer model \cite{Sch75} one
obtains the following expression
for the total cross section of a $\vphot$A collision
\beq
\label{eq:sch}
\sigma_{\vphot A}^{Sch} \left(x,Q^2,b \right) \;=\;
\frac{A\sigma_{\vphot N}}{1 \,+\, f(x, Q^2) T_A (b)} \;,
\eeq
where
\be
f(x, Q^2) = 4\pi \int_x^{\tilde{x}_\Pom} \mbox{d}x_\Pom
\,B(x_\Pom) \frac{F_{2 D}^{(3)}(x_\Pom, Q^2, \beta)}{F_2 (x, Q^2)}F_A^2
(t') ,
\ee
with $\tilde{x}_\Pom = 0.1$, where
shadowing is expected to disappear.
Here $F_2(x,Q^2)$ is the structure function for a nucleon,
$F_{2D}^{(3)}(x_\Pom,Q^2,\beta)$ is the t-integrated diffractive structure
function of the nucleon, $B(x_\Pom)$ is the $t$-slope of the diffractive
distribution, and $F_A (t')$ is the nuclear form factor where $t'
\approx - m_N^2 x_\Pom^2$. Equation~(\ref{eq:sch}) determines shadowing
for quarks (anti-quarks) in 
nuclei. 
\begin{figure}[t!]
  \centering
  \epsfig{figure=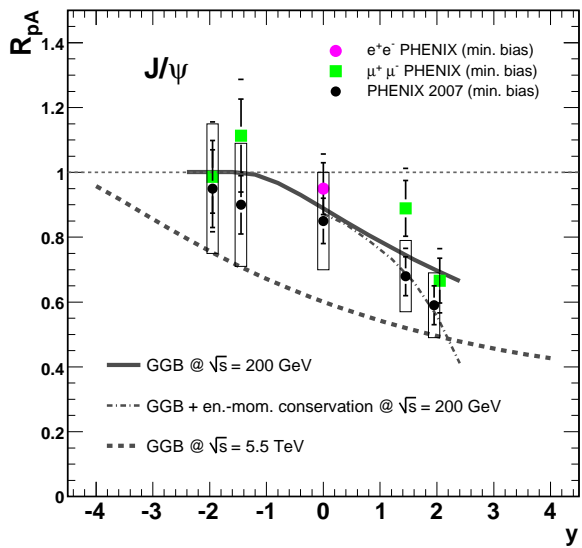,width=8.5cm}
  \caption{\label{fig:Jpsisupp_rap} Rapidity dependence of $J/\psi$
    suppression for minimum bias {\it dAu} collisions at RHIC and
    predictions for {\it pPb} collisions at LHC. Data are taken from
    \cite{PHENIX06,PHENIX07}.}
\end{figure}
For gluons the same expressions were used with substitutions 
$F_{2D}^{(3)} \left(x_\Pom,Q^2,\beta \right) \rightarrow F^g_\Pom
\left(x_\Pom,Q^2,\beta \right)$, $F_2 \left(x,Q^2 \right) \rightarrow
x g\left(x,Q^2 \right)$, indicating gluon distributions in the Pomeron,
measured in diffractive
deep inelastic scattering (DDIS), and in the proton, respectively.
Gluon distribution of
the nucleon is taken from CTEQ6M parameterization \cite{CTEQ}.
We take information
on the diffractive gluon distribution and Pomeron parameters from
recent HERA measurements \cite{H106}, where two independent fits
of the gluon diffractive distribution function, called FIT A and FIT B, represent the
overall uncertainty of extracting this information from the
measurements. We will show results only from the latter fit, denoted
as GGB, as it is closer to, as yet, preliminary combined fits where
di-jet production has been included \cite{Moz06}. This model has
previously been used to calculate quark shadowing in nucleus-nucleus
interactions in Ref.~\cite{Arm03}.

PHENIX collaboration has measured the nuclear modification factor
(NMF) of
$J/\psi$ production in {\it dAu} collisions at RHIC as a function of
centrality and rapidity in \cite{PHENIX06} and most recently in
\cite{PHENIX07} (with a more up-to-date {\it pp} reference). 
We define the centrality dependent NMF as
\beq
\label{eq:nmf}
R_{dAu} \left( \langle N_{coll} \rangle \right) \;=\;
\frac{N_{inv}^{dAu} \left( \langle N_{coll} \rangle \right)}{\langle
  N_{coll} \rangle \times N_{pp}^{inv}} \;,
\eeq
where the average number of nucleon-nucleon collisions $\langle
N_{coll} \rangle$ is obtained from the Glauber model for
a given centrality.
The results of calculations based on Eq.~(\ref{eq:sch}) are shown in
Fig.~\ref{fig:Jpsisupp} for the NMF,
given by Eq.~(\ref{eq:nmf}), at backward, mid- and forward rapidity. Since the
model of gluon shadowing does not include anti-shadowing effects, the
result is quite trivial in the backward hemisphere,
although not inconsistent with the data. Anti-shadowing is
assumed to be a 10$\%$ effect. At rapidity $y=0$ and $y=1.8$ the
consistency with experimental data is quite good.
The rapidity dependence of nuclear modification factor $R_{dAu}$ for
minimum bias {\it dAu} collisions at RHIC \cite{PHENIX06,PHENIX07} and
predictions for {\it pPb} collisions at LHC, $\sqrt{s} = $ 5.5 TeV,
are presented in Fig.~\ref{fig:Jpsisupp_rap}. At mid-rapidity, gluon
shadowing at LHC is a 40\% effect, being barely significant, $\sim10$\%, at
RHIC. This fact is important for the calculation of
charmonium yield in nucleus-nucleus collisions at these energies.

The current PHENIX data \cite{PHENIX07} indicate that at forward
rapidity shadowing alone cannot be accounted for the full drop of the
NMF of $J/\psi$. Therefore, the dash-dotted curves in
Fig.~\ref{fig:Jpsisupp} and 
Fig.~\ref{fig:Jpsisupp_rap} depict calculations including both gluon
shadowing and a model for energy-momentum conservation to be presented
in the next section. The latter effect is shown to be relevant already
at $y = 1.8$ at RHIC.

\section{Energy dependence of $\alpha(x_F)$}
\begin{figure}[t!]
  \centering
  \epsfig{figure=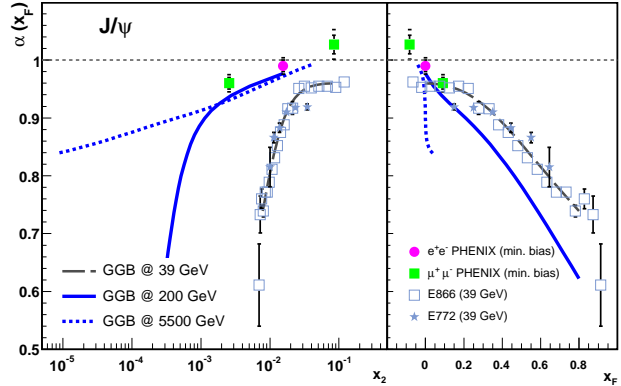,width=8.5cm}
  \caption{\label{fig:JpsiAlpha}
    $\alpha$ dependence in pA collisions vs. $x_2$
    (left) and $x_F$ (right) for $\sqrt{s} = 39-5500$ GeV. Data are
    taken from \cite{Ald91,Lei00,PHENIX06}.}
\end{figure}

Based on the previous discussion we will now formulate a model for
$J/\psi$ production in hadron-nucleus collisions at all energies.
Figure~\ref{fig:JpsiAlpha} shows experimental
points for $\alpha$ in Eq.~(\ref{eq:param}) as a function of both $x_F$ and
$x_2$ \cite{Ald91,Lei00,PHENIX06}. In order to calculate $J/\psi$
suppression at different 
energies and for all values of $x_F$ it is necessary to make a
modification of Eqs.~(\ref{eq:Abs}-\ref{eq:transition}) to describe a
transition from low-energy to high-energy regime, which happens at
$x_2^{c} \approx (m_N R_A)^{-1}$. The term $\epsilon x_+^\gamma$ in the
expression for $\xi (x_+)$ (or $\tilde{\sigma}_{\QQ}(x_+)$) provides a smooth
transition between the two regions. On the other hand, the term
$\sigma_{\QQ}(1-\epsilon)$ describing the absorption should be
modified in such a way that it should tend to zero in the high-energy
region, being substituted in this region by gluon shadowing.
To fulfill these requirements we introduce an extra multiplier
\beq
\label{eq:TransitionFunction}
f\left(x_2, x_2^{c} \right) \;=\; \exp \left\{
  -\left(x^{c}_2 / x_2 \right)^2 \right\} \;,
\eeq
for this term.
This procedure corresponds to a suppression of Glauber-type,
non-enhanced diagrams at high-energies, and has the correct
high-energy behavior, i.e., it satisfies the AGK cutting rules.

The dashed curve in Fig.~\ref{fig:JpsiAlpha} at $\sqrt{s} = $ 39 GeV has
been calculated this way with $\epsilon =$ 0.93 and $\sigma_{\QQ} =$
45 mb. Following the prescription of Eq.~(\ref{eq:transition}) for the
high-energy regime, solid and dotted curves are calculations of
gluon shadowing and energy conservation effect for RHIC and LHC
energies, respectively. The calculations for RHIC have been performed
for all $0<x_F<0.8$, while for LHC we have, for illustrative purposes,
only made calculations for the central rapidity region,
$|y|<4$. Although the coverage in $x_F$ is small at this energy,
the structure function of the nucleus is probed down to $x_2 \approx
10^{-5}$.

The model perfectly reproduces $x_F$, $x_2$ and energy
dependence of all the experimental data. Going from low to high energies we
observe a breaking of $x_F$ scaling in the central rapidity
region. The origin of this is that absorption effects,
related to longitudinally ordered rescatterings, die out while
shadowing slowly appears. This is most clearly seen in the $\alpha$
vs. $x_2$ plot.
The form of the curve for $\alpha (x_F)$ is reinstated at $x_F >$ 0.25,
although shadowing leads to a stronger overall
suppression. Additionally, scaling in $x_2$ is predicted to appear for
$J/\psi$ in the common kinematical window of RHIC and LHC, i.e. for
$10^{-3} < x_2 < 0.05$. This is a novel feature in heavy-ion
experiments and would imply the validity of the factorization theorem
in hadronic processes at ultra-relativistic energies.

Concluding, we have argued that recent data on $J/\psi$ production at
RHIC imply a profound change of the space-time picture of charmonium
production in hadron-nucleus collisions and described the experimental
data at mid-, forward and backward rapidities in terms of nuclear
shadowing. Already at $y > 1.7$ we obtain a quite strong inflence of
energy-momentum conservation, in accordance with most recent data from
experiment. 
Furthermore, we presented a model to 
describe the energy dependence of these features at $x_F$ larger or
equal zero. 
The
agreement with available data is very satisfactory. 
Nuclear
effects at $x_F < 0$ are out of the scope of this paper, and require a
separate study \cite{Boreskov03}.
These findings confirm the appearance of shadowing effects in light
particle production in {\it dAu} collisions, and will also have a
great impact on models for nucleus-nucleus collisions both at RHIC and
LHC.

\section*{Acknowledgments}
The authors would like to thank A.~Capella, C.~Pajares,
N.~Armesto and K.~Boreskov for interesting discussions. This work was
supported by the Norwegian 
Research Council (NFR) under contract No.~166727/V30,
RFBF-06-02-17912, RFBF-06-02-72041-MNTI, INTAS 05-103-7515, grant of
leading scientific schools 845.2006.2 and support of Federal
Agency on Atomic Energy of Russia.

\end{document}